\documentclass{svproc}
\usepackage{url}

\usepackage{graphicx}
\usepackage{multirow}

\begin{document}
\mainmatter              % start of a contribution
\title{Improving Neural Silent Speech Interface Models by Adversarial Training}
\titlerunning{Adversarial SSI Training}  % abbreviated title (for running head)
\author{Amin Honarmandi Shandiz\inst{1} \and László Tóth\inst{1} \and
Gábor Gosztolya\inst{2} \and \\Alexandra Markó\inst{4,5} \and Tamás Gábor Csapó\inst{3,4}}
\authorrunning{Amin Shandiz et al.} % abbreviated author list (for running head)
\tocauthor{Amin Shandiz, László Tóth, Gábor Gosztolya, Alexandra Markó,
and Tamás Gábor Csapó}
\institute{
$^1$University of Szeged, Institute of Informatics, Hungary\break
$^2$MTA-SZTE Research Group on Artificial Intelligence, Hungary\break
$^3$Budapest University of Technology and Economics, Department of Telecommunications and Media Informatics, Hungary\break
$^4$MTA-ELTE ``Lendület" Lingual Articulation Research Group, Hungary\break
$^5$Eötvös Loránd University, Dept. of Applied Linguistics and Phonetics, Hungary\break
\email{\{shandiz,tothl,ggabor\}@inf.u-szeged.hu, csapot@tmit.bme.hu, marko.alexandra@btk.elte.hu}
}

\maketitle              % typeset the title of the contribution

\begin{abstract}
Besides the well-known classification task, these days neural networks are frequently being applied to generate or transform data, such as images and audio signals. In such tasks, the conventional loss functions like the mean squared error (MSE) may not give satisfactory results. To improve the perceptual quality of the generated signals, one possibility is to increase their similarity to real signals, where the similarity is evaluated via a discriminator network. The combination of the generator and discriminator nets is called a Generative Adversarial Network (GAN). Here, we evaluate this adversarial training framework in the articulatory-to-acoustic mapping task, where the goal is to reconstruct the speech signal from a recording of the movement of articulatory organs. As the generator, we apply a 3D convolutional network that gave us good results in an earlier study. To turn it into a GAN, we extend the conventional MSE training loss with an adversarial loss component provided by a discriminator network. As for the evaluation, we report various objective speech quality metrics such as the Perceptual Evaluation of Speech Quality (PESQ), and the Mel-Cepstral Distortion (MCD). Our results indicate that the application of the adversarial training loss brings about a slight, but consistent improvement in all these metrics.
\end{abstract}
\section{Introduction}
Human speech production is a very complex motor process from speech planning in the brain to the production of the acoustic signal by precisely coordinated respiratory, laryngeal and articulatory movements~\cite{Schultz2017a}. This process has three intermediate phases which may be recorded without recording the actual speech signal~\cite{Schultz2017a,Denby2010}. In the cases of imagined speech and inner speech, there is no articulatory movement, so only the brain activity may be recorded by brain-computer interfaces like functional Magnetic Resonance Imaging (fMRI) or functional Near-Infrared Spectroscopy (fNIRS). The easiest case is when the articulatory movement is performed by the subject, as these articulatory muscle movements can be recorded by several types of devices. The basic concept of silent speech interfaces (SSI) is to reconstruct the speech signal from these articulatory signals. The most important recording methods are Electromagnetic Articulography(EMA)~\cite{Kim2017a,taguchi2018articulatory}, Ultrasound Tongue imaging (UTI)~\cite{toth20203d,Saha-ultra2speech,Hueber2010}, permanent Magnetic Articulography (PMA)~\cite{Gonzalez2017a,Fagan2008}, and surface electromyography (sEMG)~\cite{wand2018domainadversarial,Janke2017}. A video of the lip movements can also be considered as input, alone or in a combination with the above methods~\cite{wand2016lipreading,Hueber2010}. We envision the application of such SSI systems in two main situations. The first is to aid the speech communication of impaired people, such as patients after laryngectomy. The second case is when talking loudly is not appropriate or impossible because of an extreme background noise, like in certain industrial or military situations.

There are two main approaches to articulatory-to-acoustic conversion, that is, to estimate the speech signal from a recording of the articulatory movement. The two-step approach first converts the signal to text, and then converts the text to speech ~\cite{Kim2017a,Hueber2010,Fagan2008,wand2018domainadversarial}. Recently, with the invention of very powerful deep learning techniques, the direct approach has become more popular, which converts the articulatory signal to a speech signal without any intermediate steps, typically applying a deep neural network (DNN)~\cite{taguchi2018articulatory,Saha-ultra2speech,Gonzalez2017a,Janke2017}. In this paper we follow the second approach using ultrasound tongue imaging videos as the input, and we will experiment with a special DNN configuration to perform the conversion.

When we utilize DNNs to generate signals such as images or speech, the choice of the optimal loss function is not trivial, as the conventional error functions such as the mean squared error (MSE) do not fully reflect the sense of quality or similarity perceived by humans. Hence, we may expect better results by applying perceptually motivated loss functions, for which we have two options. First, speech technology has defined several objective speech quality metrics, and there were attempts to reformulate these as loss functions for DNN training~\cite{ZhangPerceptually,Martin-Donas}. Alternatively, one may train a second DNN to estimate the perceptual similarity of two images or speech signals~\cite{pihlgren2020improving,dosovitskiy2016generating}. This network is trained to discriminate `real` signals from artificially generated `fake` signals, so it will be called the discriminator, while the original network will be referred to as the generator. These two networks can be trained in parallel, and
%where the discriminator continuously improves in telling real signals from fake ones, while the generator %learns to create better and better signals that fool the discriminator. 
this combination of two nets is known as a Generative Adversarial Network (GAN). 
This work examines the applicability of GANs to improving the quality of speech signals generated by an ultrasound-based neural SSI model. We will compare the GAN-based training strategy with the conventional MSE-based training on Hungarian and English corpora, and we will evaluate the models via objective speech quality metrics.

The paper is organized as follows. In Section~\ref{sec_SSI}, we introduce the problem of generating speech in SSI systems. In Section~\ref{sec_GAN} we explain the GAN framework, and Section~\ref{sec_exp} presents the experimental conditions. The results are presented and discussed in Section~\ref{sec_res}, and the paper is closed with the conclusions in Section~\ref{sec_concl}.  

\begin{figure}[t]
\centering
\includegraphics[width=1.0\textwidth]{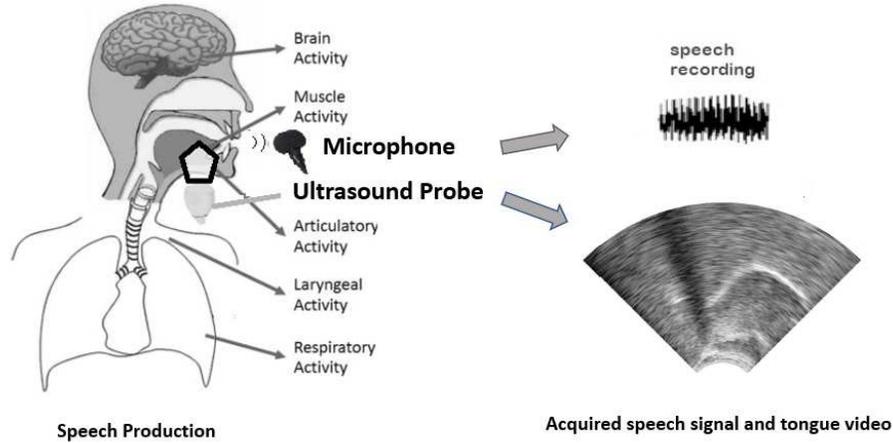}
\caption{Illustration of the data acquisition process.} \label{fig_SSI}
\end{figure}

\section{Silent Speech Interfaces Built on Ultrasound Tongue Imaging and Neural Vocoders}
\label{sec_SSI}
Here, we worked with an ultrasound-based data acquisition technique. In this framework, an ultrasound probe is placed under the chin of the subject, so it can record images of the tongue movement in a midsagittal orientation (see Fig~\ref{fig_SSI}). The device applied was the "Micro" ultrasound system produced by Articulate Instruments Ltd. To support DNN training, the subjects were asked to speak loud, and their speech was recorded in parallel with the ultrasound output (cf. Fig~\ref{fig_SSI} again). The ultrasound and speech recordings served as the training inputs and training targets for the DNN, respectively. As the input and output signals were synchronized, we applied conventional network structures which perform the conversion in a time-synchronized, frame-by-frame manner.  

Training a network to generate speech signals directly would require a huge amount of training data. Hence, we opted for a two-step solution, which was motivated by speech synthesis. In text-to-speech systems, a popular approach is to first convert the text to a spectrogram, then convert the spectrogram to a speech signal. Nowadays, both steps can be implemented by neural networks~\cite{waveglow}. To adjust this approach to our task, only the first network needs to be modified, as our input is an ultrasound video and not a text. This way, we can apply large, pre-trained networks for the second task, while our network has to estimate only a dense spectral representation instead of the speech waveform itself. Several neural vocoders are available for the speech generation task~\cite{govalkar}, and we chose to use the WaveGlow model~\cite{waveglow}, as we applied it successfully earlier~\cite{Csapo2020}. As WaveGlow requires a sequence of 80-dimensional mel-scaled spectral vectors as input, the task of our network was to estimate such a spectral vector form each frame of the video. The next section discusses the details of this step.

\section{Generative Adversarial Networks for Articulatory-to-Acoustic Mapping}
\label{sec_GAN}
As we use WaveGlow for the speech synthesis step, the task for our DNN was to convert each ultrasound video frame (64x128 pixels) into a mel-spectral vector (with 80 components). In an earlier paper we found that the best results can be achieved by using a short sequence of video frames as input, rather than just a single frame~\cite{toth20203d}. We got the lowest error rate with a special 3D convolutional neural network (CNN) structure, and other authors reported similar findings~\cite{Saha-ultra2speech}. Here, we apply the same 3D CNN structure that was found the best in our earlier study~\cite{toth20203d}. The input of this network consists of a 25-frame block of the ultrasound video, and the output is a mel-spectral vector. For more details on this network, we refer the reader to our earlier study. 

A crucial parameter of DNN training is the loss function which formalizes the difference between the training targets and the DNN output. In regression tasks, the simplest solution is to use general loss functions such as the mean-squared error (MSE). However, such mathematically motivated loss functions may not conform with our perception. When the network output is a speech signal, such as in speech enhancement, one may try to apply objective speech quality metrics as the loss function~\cite{ZhangPerceptually,Martin-Donas}. Here, we explore another method where a second, discriminator network is trained to judge the quality of the output signal. This approach leads to the subject of Generative Adversarial Networks (GANs).

\begin{figure}[t]
\centering
\includegraphics[width=1.0\textwidth]{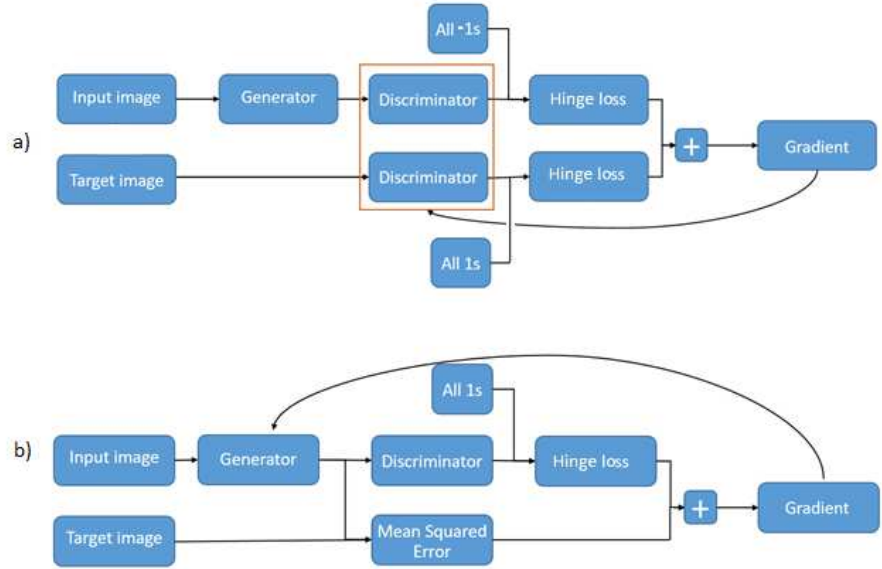}
\caption{The error calculation (forward arrows) and weight update (backward arrows) training steps for the discriminator (upper image) and the generator (lower image) networks of the GAN.} \label{fig_GAN}
\vspace{-1mm}
\end{figure}

GANs were first proposed by Goodfellow et al~\cite{goodfellow2014generative}. A GAN is basically a combination of two networks, a generator and a discriminator. The goal of the generator is to artificially generate data samples as if they were coming from a real distribution, for example, as if they were real images. The distribution to be approximated is not mathematically defined, but it is estimated by the other network, the discriminator. For this purpose, the discriminator is trained to separate real data vectors from fake ones (ie. those created by the generator), using real and fake data items. The generator and the discriminator are normally trained in parallel, in a sort of minimax game: the discriminator seeks to improve in differentiating real data items from fake ones, so its error is minimized, while the generator tries to generate fake images that are very similar to the real ones, so it seeks to increase the error of the discriminator. This system is called adversarial, as the generator and the discriminator work against each other.

When the goal is not data generation but data transformation, we can modify the original GAN formulas by adding the input to be transformed as a `condition` of the data distribution, and hence this model is called the conditional GAN or CGAN~\cite{mirza2014conditional}. GANs and CGANs were shows to produce very good quality images in broad types of vision tasks like image generation~\cite{goodfellow2014generative}, image translation~\cite{isola2017image,zhu2017unpaired}, and text-to-image synthesis~\cite{reed2016generative}. In speech technology, the GAN-based approach is the most successful in speech enhancement~\cite{zhang2020loss} and voice conversion~\cite{kaneko19}. 

 In our case, the role of the generator was played by the network that converts the ultrasound data to mel-spectral data. For this purpose we used the same network as in our previous study~\cite{toth20203d}. As the baseline system, we trained this network conventionally, using the MSE loss function. Then we created a discriminator which was trained to discriminate real mel-spectrograms from those produced by the generator. As was explained earlier, the generator was trained in an adversarial manner, so it was optimized to create spectrograms that cannot be discriminated from real spectrograms by the discriminator.
 
 The generator and the discriminator was trained in parallel, using a two-step process~\cite{zhang2020loss}. As shown in Fig.~\ref{fig_GAN}a, the first step trains the discriminator on two types of images: on real spectrograms (lower data path), and on spectrograms created by the generator (upper path). These serve as positive and negative training examples, denoted by the target labels of 1 and -1 in the figure. For the training, we applied the hinge loss function, which is frequently used as a GAN objective~\cite{Melgan}, and optimized it using the Adam optimizer with a learning rate of 0.0002. In this step, only the weights of the discriminator are updated, while the generator weights are frozen.
 
 Fig.~\ref{fig_GAN}b shows the other training step that updates the generator weights (now the discriminator weights are frozen). We combine two error functions to calculate the loss, and hence the gradient. First, we compare the generator output and the target spectrograms using the MSE loss (lower path). The second loss value is obtained from the discriminator (upper path). Notice that now the discriminator target labels are flipped, as we wish to train the generator in an adversarial manner, to create outputs that look like a real spectrogram.
\section{Experimental Set-Up}
\label{sec_exp}

The goal of the experiments was to compare the performance of the 3D CNN generator network trained conventionally, using the MSE loss function, with the training scheme that applies the GAN-style discriminator network. In the experiments we evaluated our models on two data sets -- one of them being recorded from a Hungarian speaker and the other from an English speaker. Here, we shall present the technical details of the experiments.
\subsection{Data Sets and Data Preprocessing}

{\bf Hungarian Data Set:} The parallel ultrasound and speech recordings were collected from a Hungarian female subject reading sentences aloud, using the equipment briefly described in Section~\ref{sec_SSI}. The whole duration of the recordings was about half an hour (438 sentences), from which 310 were used for training, 41 for development and 87 for testing, respectively.

The ultrasound transducer produces an ultrasound video of the tongue movement at a rate of 82 frames per second. One frame of this video has a resolution of 64x946; that is, the device collects 946 data samples along 64 scan lines. As these ultrasound images are very noisy and they contain very few details, we decreased the image size to 64x128 by applying a bicubic interpolation. The pixel intensities were min-max scaled to the [-1, 1] range.

The speech signals were recorded in parallel with the ultrasound video at a sampling rate of 22050 Hz. The speech and ultrasound signals were synchronized using the software tool provided by Articulate Instruments.

{\bf English Data Set:} As the English data set, we used the TAL corpus~\cite{ribeiro2020tal}. It contains parallel speech, tongue ultrasound and lip video recordings from 81 speakers. Here, we just used the TAL1 subset of the corpus, which contains the recordings of a single trained native English speaker. The recording conditions were very similar to that of the Hungarian data set and, after division, the train, validation and test sets contained 1015, 50 and 24 utterances, respectively. We applied exactly the same preprocessing steps as for the Hungarian data. 

\begin{table}[t]
\caption{The layers of the discriminator network (following Keras' terminology). Albeit not shown, each but the last Conv2D layer is followed by batch normalization.}
\centering
\renewcommand{\arraystretch}{1.1} 
\begin{tabular}{|l|c|c|c|c|c|}
\hline
Layer & ~Filters~ & ~~Size~~ & ~Strides~ & ~Padding~ & ~Activation~~\\
\hline
\hline
Conv2D&64& (4,4)& (2,2)& same&relu\\
%BatchNorm &--&--&--&--&--\\
Conv2D&128& (4,4)& (2,2)& same&relu\\
%BatchNorm &--&--&--&--&--\\
Conv2D&256& (4,4)& (2,2)& same&relu\\
%BatchNorm &--&--&--&--&--\\
ZeroPadding2D~~&--&--&--&--&--\\
Conv2D&512& (2,2)&(1,1))&valid&relu\\
%BatchNorm &--&--&--&--&--\\
ZeroPadding2D&--&--&--&--&--\\
Conv2D&1&(4,4)&(1,1)&valid&tanh\\
\hline
\end{tabular}
\label{table_DISC}
\end{table}

\subsection{DNN Configuration and Training}

As the generator we used the 3D CNN from our earlier study~\cite{toth20203d}, with the slight modification that instead of one target vector, here we specified 5 consecutive spectral vectors as targets. We hoped this modification would improve the performance of the discriminator.

As the discriminator, we applied the so-called Patch-GAN method~\cite{isola2017image,zhu2017unpaired}. Instead of a two-class decision for the whole image, the discriminator of a patch-GAN returns a set of output values for different regions of the image. We applied a fully convolutional CNN for this purpose, with an output vector of 10 components. The actual network configuration and its parameters are shown in Table~\ref{table_DISC}. 

When synchronized training targets are available, as in our case, adversarial training can be combined with conventional MSE-training. We got the best results when the MSE loss function was combined with the adversarial training loss using a weighting ratio of 0.75-0.25 for the two loss functions, respectively.

\begin{table}[t]
\caption{The MSE and $R^2$ scores of the generator with the MSE and GAN training approaches.}
\centering
\renewcommand{\arraystretch}{1.2} 
\begin{tabular}{|c||c|c|c|c||c|c|c|c|}
\hline
 & \multicolumn{4}{c||}{Hungarian Corpus}                & \multicolumn{4}{c|}{English Corpus}                  \\ \cline{2-9} 
                                                ~Training~ & \multicolumn{2}{c|}{Dev} & \multicolumn{2}{c||}{Test} & \multicolumn{2}{c|}{Dev} & \multicolumn{2}{c|}{Test} \\ \cline{2-9} 
                                                                        Method & ~MSE~        & ~Mean  $R^2$~      & ~MSE~         & ~Mean  $R^2$~      & ~MSE~        & ~Mean  $R^2$~     & ~MSE~         & ~Mean  $R^2$~      \\ \hline
MSE                                                             & 0.327      & 0.68        & 0.326       & 0.677       & 0.316      & 0.683       & 0.317       & 0.68        \\ \hline
GAN                                                                      & 0.287      & 0.719       & 0.29        & 0.713       & 0.287      & 0.715       & 0.293       & 0.71        \\ \hline
\end{tabular}
\label{table_Eres}
\end{table}

\section{Results and Discussion}
\label{sec_res}

Evaluating the quality of the generator is not trivial, as its output is a spectrogram, which is converted
to a speech signal by WaveGlow. Our main goal is to increase the quality of the synthesized speech, and we have two ways to measure this quality. First, we can perform subjective listening tests such as MUSHRA~\cite{mushra}. Unfortunately, this would require a lot of human subjects, hence it would be slow and troublesome, which we wished to avoid here. Instead, we can evaluate objective, formally defined metrics. As the simplest of these, we report two metrics that are popular in machine learning for regression tasks, namely the mean squared error and the correlation-based mean $R^2$ score (the former has to be minimized, while the latter should be maximized). As can be seen in Table~\ref{table_Eres}, all these metrics gave a slight improvement when the GAN-style training was applied, both for the Hungarian and English subjects, and both for the development and the test sets.

As we mentioned in Section~\ref{fig_SSI}, our output is a speech signal, and in this case the above simple metrics may not perfectly reflect human perception. Fortunately, several objective measures have been developed in speech technology and in telecommunications to compare the quality or intelligibility of speech recordings. These measures include the Short-Time Objective Intelligibility (STOI)~\cite{Martin-Donas} and its extended version (ESTOI),
the perceptual evaluation of speech quality (PESQ) method~\cite{Martin-Donas}, and its extended version known as perceptual metric for speech quality evaluation (PMSQE)~\cite{Martin-Donas}. While the first three measure quality, and hence a higher value means a better performance, PMSQE was designed to be applicable as a loss function in DNN training, so in this case a lower value indicates better quality. We also evaluated the signal-to-distortion ratio (SDR) and its extended, scale invariant version (SI-SDR)~\cite{SDR}. As their name suggests, for these metrics a higher value means better quality. Finally, we calculated the mel-cepstral distortion (MCD)~\cite{Martin-Donas}, which is a distortion metric, so it should be minimized to increase speech quality. These results are summarized in Table~\ref{table_Tres} for the test sets of the two databases. To aid readibility, the two metrics that are to be minimized were placed in the rightmost columns, and the best scores are highlighted for both corpora. As the results indicate, extending the MSE training criterion with GAN-style adversarial training led to a consistent improvement in all the evaluated metrics and for both corpora. Although in certain cases the improvement is only slight, the results clearly justify the utility of generative adversarial networks for the articulatory-to-acoustics mapping task. In comparison, Ribeiro et al. reported an MCD score of 2.99 for the English dataset using a much more sophisticated endocer-decoder neural architecture~\cite{ribeiro2020tal}.

\begin{table}[t!]
\caption{Objective quality scores for the two training methods and the two corpora.}
\centering
\renewcommand{\arraystretch}{1.2} 
\begin{tabular}{|c|c|c|c|c|c|c|c|}
\hline
\begin{tabular}[c]{@{}c@{}}Corpus and \\ Training method\end{tabular} & ~\textbf{STOI}~   & ~\textbf{ESTOI}~  & ~\textbf{PESQ}~  & ~\textbf{SISDR}~   & ~\textbf{SDR}~     & ~\textbf{PMSQE}~ & ~\textbf{MCD}~   \\ \hline\hline
Hun - MSE                                        & 0.7050          & 0.456           & 1.282          & -39.363          & -18.021          & 2.797          & 4.627          \\ \hline
~Hun - GAN~                                                        & \textbf{0.7067} & \textbf{0.4673} & \textbf{1.311} & \textbf{-37.868} & \textbf{-16.893} & \textbf{2.777} & \textbf{4.558} \\ \hline\hline
Eng - MSE                                              & 0.612           & 0.385           & 1.464          & -36.176          & -18.31           & 2.827          & 3.38           \\ \hline
Eng - GAN                                                          & \textbf{0.623}  & \textbf{0.405}  & \textbf{1.503} & \textbf{-36.031} & \textbf{-17.672} & \textbf{2.735} & \textbf{3.229} \\ \hline
\end{tabular}
\label{table_Tres}
\end{table}

\section{Conclusions}
\label{sec_concl}

The application of the GAN framework has already proved successful in speech enhancement and voice conversion tasks, and here we made the first attempts to apply it to the articulatory-to-acoustic mapping task of ultrasound-based silent speech interfaces. 
As the baseline, we trained the generator network conventionally, using the MSE loss, and then we extended its training with adversarial training by means of a discriminator. We applied our method on two data sets, a Hungarian corpus and an English corpus, and in both cases we found that the quality of the generated speech signals improved, according to several objective speech quality metrics. In the future, we plan to extend our framework to the direct generation of speech, such as by using MelGAN~\cite{Melgan} , and we also plan to study the effect of incorporating perceptual speech quality metrics into the training loss~\cite{Martin-Donas}.

\section{Acknowledgments}

This study was supported by the grant NKFIH-1279-2/2020 of the Ministry for Innovation and Technology, Hungary, and by the Ministry of Innovation and the National Research, Development and Innovation Office through project FK 124584 and within the framework of the Artificial Intelligence National Laboratory Programme. Gábor Gosztolya was supported by the UNKP 20-5 National Excellence Programme of the Ministry of Innovation and Technology, and by the J\'anos Bolyai Research Scholarship of the Hungarian Academy of Science. The GPU card used for the computations was donated by the NVIDIA Corporation. 
%\begin{thebibliography}{6}
%
%\bibitem {smit:wat}
%Smith, T.F., Waterman, M.S.: Identification of common molecular subsequences.
%J. Mol. Biol. 147, 195?197 (1981). \url{doi:10.1016/0022-2836(81)90087-5}
%\end{thebibliography}

%\bibliographystyle{splncs04}
\bibliographystyle{splncs04_unsrt}
\bibliography{AICV2021}

\end{document}